\documentclass{ws-ijmpa}
\usepackage{url}
\usepackage[super,compress]{cite}
\usepackage{graphicx}
\usepackage{hyperref}
\hypersetup{colorlinks,urlcolor=black,citecolor=black,linkcolor=black,filecolor=black}
\usepackage{breakurl}
\begin{document}
\markboth{Denis Bodrov}{Tau physics at Belle and Belle~II}

%
\catchline{}{}{}{}{}
%

\title{Tau physics at Belle and Belle~II}

\author{Denis Bodrov \\
(on behalf of the Belle and Belle II Collaborations)
}

\address{Soochow University, \\
1 Shizi Street, 
Suzhou 215006, P. R. China\\
bddbodrov@stu.suda.edu.cn}

\address{National Research University Higher School of Economics, \\
20 Myasnitskaya Street, Moscow 101000, Russia\\
dbodrov@hse.ru}

\maketitle

\begin{history}
\received{Day Month Year}
\revised{Day Month Year}
\end{history}

\begin{abstract}
We present here the major results obtained in $\tau$ physics by the Belle and Belle~II experiments. For the Belle~II experiment, we also discuss prospects for improved measurements of the $\tau$ lepton properties, new results in Michel parameters determination, and searches for $CP$ and lepton flavor violation in $\tau$ decays.

\end{abstract}

\ccode{PACS numbers: 13.35.Dx, 12.15.Ji, 14.60.Fg}


\section{Introduction}	

The $\tau$ lepton is the heaviest known lepton with a mass of around $1.777$ GeV$/c^2$ allowing for both leptonic and hadronic decay modes. 
Due to its larger mass compared to the muon, the $\tau$ lepton may be more sensitive to some models of new physics (NP)~\cite{Krawczyk:2004na,Chun:2016hzs,Marquez:2022bpg}.

The list of all possible measurements in $\tau$ physics is wide. It includes the precise determination of its parameters: mass, lifetime, and electric and magnetic dipole moments (EDM and MDM). The study of pure leptonic decays allows for testing lepton flavor universality and determination of the Lorentz structure of weak interaction without uncertainties of QCD calculations. The existence of hadronic decay modes opens up an opportunity to study QCD at 1 GeV, test lepton flavor universality, and search for $CP$ violation (CPV). Any deviations from the Standard Model (SM) predictions observed in these studies will indicate the presence of the new physics. In addition, one can directly search for NP in lepton-flavor-violating (LFV) decays of $\tau$ lepton or in decays with new invisible particles in the final state.

\section{Experiment}

In $\tau$ physics, experiments located at $e^+e^-$ colliders, in general, outperform experiments located at hadron machines because the initial $\tau^+\tau^-$-pair state is known, and the detectors are nearly hermetic; therefore, neutrinos in the final state are the only undetected particles in the majority of events. 

Existing BESIII~\cite{BESIII:2009fln} and KEDR~\cite{Anashin:2002uj} experiments are limited in statistics, while $B$-factories, mainly operated at a center-of-mass (c.m.) energy of 10.58 GeV, are perfect facilities to study $\tau$ lepton due to unprecedented $\tau^+\tau^-$-pair samples collected by them. 

\subsection{Belle experiment}
The Belle detector~\cite{Belle:2000cnh, Belle:2012iwr} is located at the KEKB asymmetric-energy $e^+e^-$ collider~\cite{Kurokawa:2001nw, Abe:2013kxa}. It is a large-solid-angle magnetic spectrometer that consists of a silicon vertex detector (SVD), a 50-layer central drift chamber (CDC), an array of aerogel threshold Cherenkov counters (ACC), a barrel-like arrangement of time-of-flight scintillation counters (TOF), and an electromagnetic calorimeter (ECL) composed of CsI(Tl) crystals located inside a superconducting solenoid coil that provides a 1.5~T magnetic field. An iron flux return located outside of the coil is instrumented to detect $K_L^0$ mesons and to identify muons (KLM). The integrated luminosity of $988\,\text{fb}^{-1}$ collected by the Belle detector corresponds to approximately $912\times 10^6$ $\tau^+\tau^-$ pairs.

\subsection{Belle II experiment}
The Belle II experiment is a major upgrade of its ancestor Belle experiment. It is located at the SuperKEKB asymmetric-energy $e^+e^-$ collider~\cite{Akai:2018mbz}.

The Belle II detector consists of subsystems arranged cylindrically around the interaction region~\cite{Belle-II:2010dht, Kou:2018nap}. Belle II uses cylindrical coordinates in which the $z$-axis is approximately collinear with the electron beam. Charged-particle trajectories (tracks) are reconstructed by a two-layer silicon-pixel detector (PXD) surrounded by a four-layer double-sided silicon-strip detector and a central 56-layer drift chamber. The latter two detectors also measure the ionization energy loss. A quartz-based Cherenkov counter measures both the direction and time-of-propagation of photons and identifies charged hadrons in the central region, and an aerogel-based ring-imaging Cherenkov counter identifies charged hadrons in the forward region. An electromagnetic calorimeter made of CsI(Tl) crystals measures photon and electron energies and directions. The above subdetectors are immersed in a 1.5 T axial magnetic field provided by a superconducting solenoid. A subdetector dedicated to identifying muons and $K_L^0$ mesons is installed outside of the solenoid.

By the end of the operation of the Belle~II experiment, it is planned to collect an integrated luminosity of $50\,\text{ab}^{-1}$ corresponding to $46\times 10^9$ $\tau^+\tau^-$ pairs. 

In addition to the improved machinery, a new Neural-Network-based low multiplicity trigger was developed at Belle~II, significantly improving the analysis efficiency for the $\tau$ physics compared to Belle.

\section{Measurement of the \boldmath{$\tau$} lepton properties}
\subsection{Mass}
The mass of the $\tau$ lepton is one of the fundamental parameters of the SM that is obtained experimentally. Its precise knowledge is required for the lepton universality tests~\cite{HFLAV:2022esi} and measurement of the strong interaction coupling at the $\tau$ mass scale $\alpha_s(m_\tau)$~\cite{Davier:2005xq,Davier:2013sfa}.

There are two ways to determine the mass of the $\tau$ lepton: measure the $e^+e^-\to\tau^+\tau^-$ cross-section at its threshold (done by KEDR~\cite{KEDR:2006xae} and BESIII~\cite{BESIII:2014srs} experiments) or use the pseudomass method in $\tau^-\to\pi^-\pi^+\pi^-\nu_\tau$\footnote{Charge conjugation is implied throughout the paper unless otherwise indicated.} decay~\cite{ARGUS:1992chv} (used in the experiments at $e^+e^-$ colliders working above $\tau^+\tau^-$-pair threshold). With the latter method, we measure the mass of the $\tau$ lepton using the integrated luminosity of 190 fb$^{-1}$ collected by the Belle~II detector.

\begin{figure}[h]
\centerline{\includegraphics[width=0.5\textwidth]{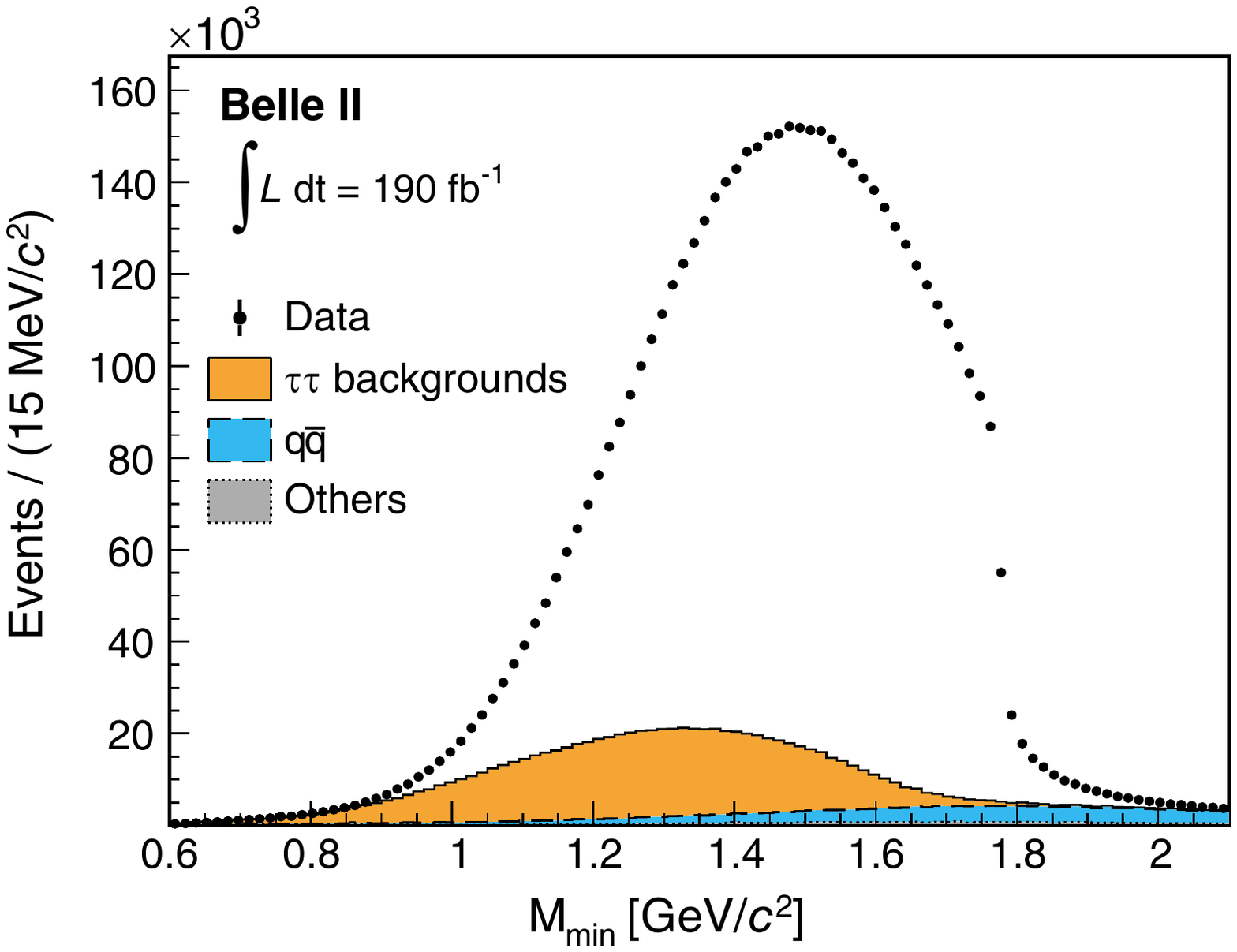}}
\caption{Spectrum of $M_\text{min}$ in experimental data (dots) with simulated background (filled histograms). \label{fig:1}}
\end{figure}

The pseudomass is defined as follows:
\begin{equation} \label{eq:1}
    M_\text{min} = \sqrt{M_{3\pi}^2 + 2(\sqrt{s}/2-E_{3\pi}^*)(E_{3\pi}^*-p_{3\pi}^*)}\leq m_\tau,
\end{equation}
where the asterisk refers to the variables in the c.m. frame, and we used the fact that $E_\tau^*$ is equal to half of the beam energy $\sqrt{s}$, neglecting the initial-state radiation. In the ideal situation, the sharp edge of the $M_\text{min}$ distribution would extend up to the mass of the $\tau$ lepton. However, in the real experiment, the endpoint is smeared by the resolution, initial- and final-state radiation, and background, as it is shown in Fig.~\ref{fig:1}. From the fit of the edge, we extract the $\tau$ lepton mass. 

\begin{figure}[h]
\centerline{\includegraphics[width=1\textwidth]{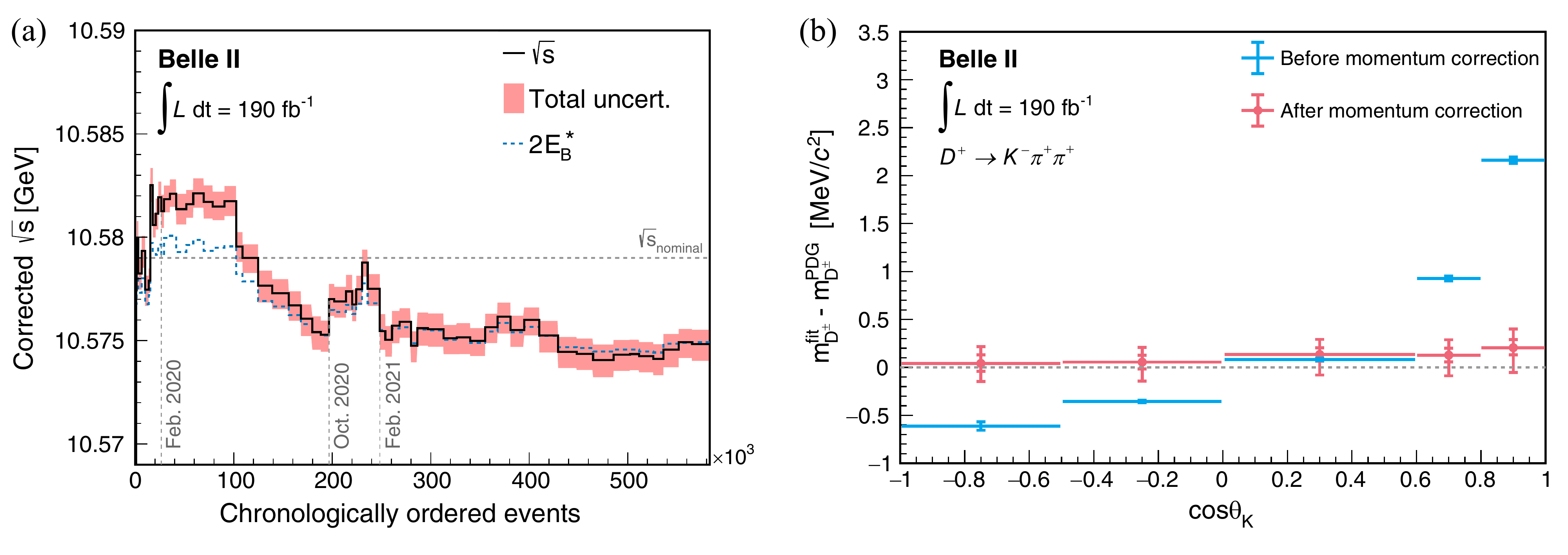}}
\caption{(a) Corrected c.m. energy $\sqrt{s}$ (solid line) and
c.m. energy of $B\bar{B}$ pair $2E_B^*$ (dashed blue line) as functions of data-taking time, expressed in terms of chronologically ordered event numbers. (b) Deviation of the $D^+$ invariant-mass peak position from the known value before (blue) and after (red) momentum corrections as a function of the cosine of the kaon polar angle $\theta_K$. \label{fig:2}}
\end{figure}

The precision of the most recent measurements is dominated by systematic uncertainty, which control is also the main challenge in this measurement and requires an excellent understanding of the detector performance and the backgrounds. The main sources of systematics are the calibration of the beam energy and the determination of the daughter pion momenta, as can be seen from Eq.~\eqref{eq:1}. The former one was performed using $B\bar{B}$-pair production cross-section and its hadronic decays. The corrected beam energy is shown in Fig.~\ref{fig:2}(a). The scale factors for the daughter pion momenta were obtained from the $D^0\to K^-\pi^+$ sample with cross-checks in the $D^+\to K^-\pi^+\pi^+$, $D^0\to K^-\pi^+\pi^-\pi^+$, and $J/\psi\to\mu^+\mu^-$ samples. The difference between the reconstructed and nominal masses of the $D^+$ meson before and after corrections is shown in Fig.~\ref{fig:2}(b).

The systematic uncertainty of the beam energy calibration comes mainly from the uncertainties in the energy dependence of the $e^+e^-\to B\bar{B}$ cross-section~\cite{BaBar:2008cmq,Belle:2021lzm} and the average values of the $B^0$ and $B^+$ meson masses~\cite{Workman:2022ynf}. This systematic uncertainty is estimated to be $0.07$ MeV$/c^2$. The charged-particle momentum correction leads to an additional $0.06$ MeV$/c^2$ uncertainty. Combination with other sources leads to the total systematic uncertainty of $0.11$ MeV$/c^2$. 

Finally, Belle~II provides the most precise measurement of the $\tau$ lepton mass: $1777.09\pm0.08\pm0.11$ MeV$/c^2$, where the first uncertainty is statistical, and the second one is systematic~\cite{Belle-II:2023izd}.

\subsection{Lifetime}
The $\tau$ lepton lifetime is another parameter, precise knowledge of which is required for the lepton universality test~\cite{HFLAV:2022esi}. For its measurement, a boost of the $\tau$ lepton in the laboratory frame is needed. $B$-factories being developed for the precise determination of the time-dependent $CP$ violation are perfectly suitable for such studies. Current world's best result of $[290.17\pm0.53\pm0.33]\times10^{-15}$ s, where the first uncertainty is statistical, and the second one is systematic, is obtained using the integrated luminosity of 711 fb$^{-1}$ collected by the Belle detector. In addition, a $CPT$ test is provided: $(\langle \tau_{\tau^+} \rangle - \langle \tau_{\tau^-} \rangle)/ \langle \tau_{\tau} \rangle < 7.0\times10^{-3}$ (90\% CL), where the largest contribution to the systematic uncertainty cancels in the difference, and the final precision is determined by statistics.

\begin{figure}[h]
\centerline{\includegraphics[width=0.5\textwidth]{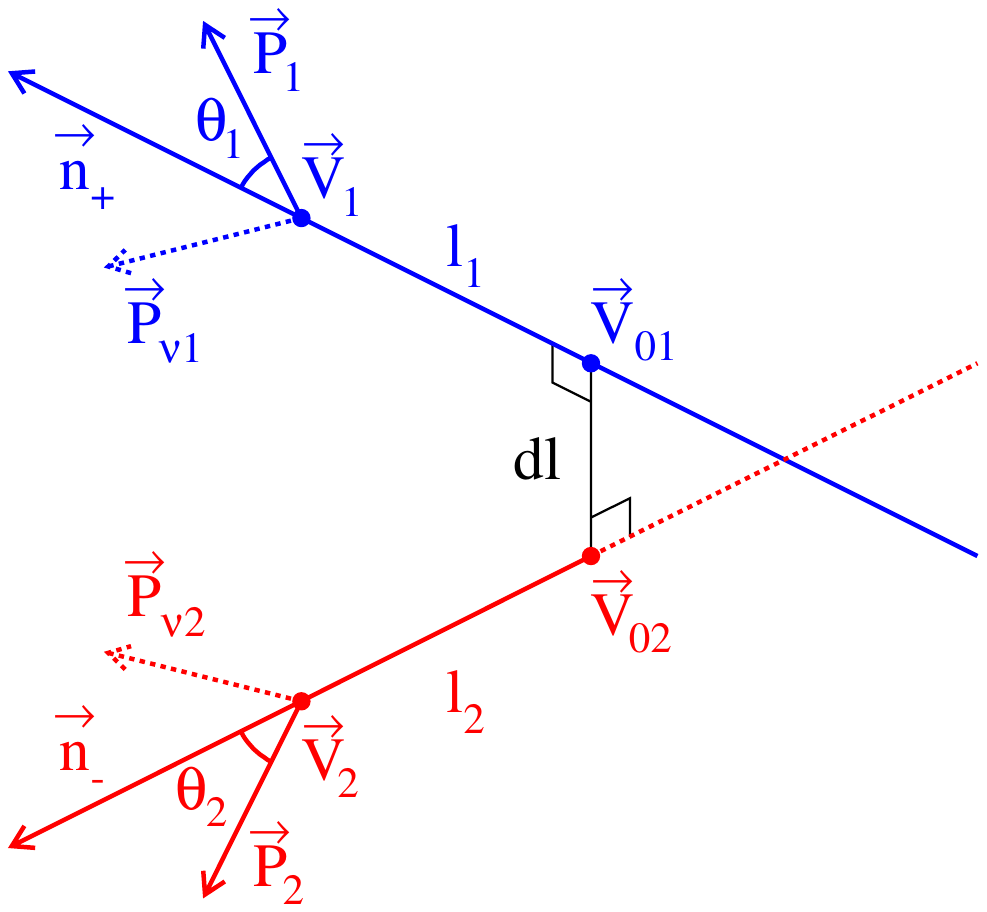}}
\caption{The schematic view of the $\tau^+\tau^-$ event in the laboratory frame. $\vec{P}_1$ and $\vec{P}_2$ are the combined momenta of the three hadron systems for each $\tau$ decay. \label{fig:3}}
\end{figure}

To measure the lifetime, we reconstruct both $\tau$ leptons in the event decaying into three pions. Three charged particles in the final state allow for precise determination of the $\tau$ lepton decay vertex. To reconstruct the production vertex, we find the point of the closest approach of momentum directions of the $\tau^+\tau^-$ pair. Due to the presence of neutrinos in the final state, the hadronic decay mode can provide $\tau$ lepton momentum direction with twofold ambiguity only~\cite{Kuhn:1993ra}. We take the mean value of two vectors as the momentum direction. The general scheme to illustrate this procedure is shown in Fig.~\ref{fig:3}.

The result precision obtained by the Belle experiment is limited by the statistical uncertainty, and the main systematics arise from the SVD alignment. In future, the Belle~II experiment with larger statistics and improved vertex detector, already showing two times better resolution in $D$-meson lifetime measurement~\cite{Belle-II:2021cxx}, can reproduce this study with higher precision. 

The signal efficiency of the described approach can be increased at Belle~II by replacing the three-pion decay mode of the tagging $\tau$ lepton with $\tau^+\to\rho^+\bar{\nu}_\tau$, which has more than two times higher branching fraction. Thanks to the SuperKEKB nanobeam collision scheme, we can reconstruct the production point of the $\tau^+\tau^-$ pair by applying a constraint to the beam spot.

\subsection{EDM and MDM}
The general expression of the $\gamma\tau\tau$ vertex includes Electric and Magnetic Dipole Moments. In the SM, the first one is almost forbidden by the $T$ invariance, and the second one has a value of $a^\text{SM}_\tau=117721(5) \times 10^{-8}$. 

The most precise measurement of EDM is done by the Belle experiment using the integrated luminosity of 833 fb$^{-1}$~\cite{Belle:2021ybo}. The applied method exploits triple momentum and spin correlation observables (so-called optimal observables) built from the matrix element
\begin{equation}
 M^2=M^2_{\text{SM}}+\Re(d_\tau)M^2_{\Re}+\Im(d_\tau)M^2_{\Im}+|d_\tau|^2M^2_{d^2}
\end{equation}
as $O_\Re={M^2_\Re}/{M^2_\text{SM}}$ and $O_\Im={M^2_\Im}/{M^2_\text{SM}}$. The averages of these observables depend linearly on the real and imaginary parts of EDM: $\langle O_\Re\rangle=a_\Re\Re(d_\tau)+b_\Re$ and $\langle O_\Im\rangle=a_\Im\Im(d_\tau)+b_\Im$. The obtained boundaries are $-1.85\cdot 10^{-17}<\Re(d_\tau)<6.1\cdot 10^{-18}  \,e\text{cm}$ and $-1.03\cdot 10^{-17}<\Im(d_\tau)<2.3\cdot 10^{-18}  \,e\text{cm}$ (95\% CL). Using full integrated luminosity Belle~II can improve this result up to $|\Re,\Im(d_\tau)|<(0.1$--$1)\times10^{-18}\,e\text{cm}$~\cite{Belle-II:2022cgf}.

Currently, MDM was measured once by the DELPHI collaboration in two-photon interaction, and only an upper limit was set~\cite{DELPHI:2003nah}. Using the same approach of the optimal observables as in the EDM measurement, Belle~II can achieve sensitivity to the NP contribution at the level of $2\times10^{-5}$ with the full integrated luminosity~\cite{Chen:2018cxt}.

\section{Study of pure leptonic $\tau$ decays}
\subsection{Lepton Flavor Universality}
Leptonic $\tau$ decays allow for precise testing of the $e$--$\mu$ lepton universality:
\begin{equation}
    \left(\dfrac{g_\mu}{g_e} \right)_\tau=\sqrt{R_\mu\dfrac{f(m_e^2/m_\tau^2)}{f(m_\mu^2/m_\tau^2)}},
\end{equation}
where
\begin{equation}
    R_\mu = \dfrac{\mathcal{B}(\tau^-\to\mu^-\bar{\nu}_\mu\nu_\tau(\gamma))}{\mathcal{B}(\tau^-\to e^-\bar{\nu}_e\nu_\tau(\gamma))}\overset{\text{SM}}{=}0.9726, 
\end{equation}
and 
\begin{equation}
    f(x) = 1 - 8x + 8x^3 - x^4 - 12x^2\ln{x}.
\end{equation}

Using integrated luminosity of 362 fb$^{-1}$ collected by the Belle~II detector, we provide a result of $R_\mu=0.9675\pm0.0007\pm 0.0036$, where the first uncertainty is statistical, and the second one is systematic, leading to $|g_\mu/g_e|=0.9974\pm0.0019$~\cite{Belle-II:2024vvr}. This is the most precise test in a single measurement. Here, the systematic uncertainty is larger than the statistical one, with the leading contribution coming from particle identification (0.32\%) and trigger efficiency (0.10\%).

\subsection{Michel parameters}
Leptonic $\tau$ decays can also be used to determine Lorentz structure of the charged currents interaction in the theory of weak interaction by measurement of the so-called Michel parameters~\cite{Michel:1949qe}. They are bilinear combinations of the coupling constants $g^\gamma_{\varepsilon\mu}$ arising in the most general expression for the decay matrix element~\cite{Scheck:1984md,Mursula:1984zb,Fetscher:1986uj}:
\begin{equation}
    M=\dfrac{4G_F}{\sqrt{2}}\sum_{\substack{\gamma=S,V,T\\
\varepsilon,\mu=R,L} }g^\gamma_{\varepsilon\mu}\langle \bar \ell_\varepsilon \vert \Gamma^\gamma \vert ((\nu_\ell)_\alpha)\rangle\langle (\bar \nu_\tau)_\beta
 \vert \Gamma_\gamma \vert \tau_\mu\rangle,
\end{equation}
where
\begin{eqnarray} 
\Gamma^S=1, \quad\Gamma^V=\gamma^\mu, \text{ and } \Gamma^T=\dfrac{i}{2\sqrt{2}}(\gamma^\mu\gamma^\nu-\gamma^\nu\gamma^\mu),
\end{eqnarray}
and $\varepsilon,\mu=L,R$ are left- and right-handed leptons, respectively. In the SM, the only nonzero term is $g^V_{LL}=1$. Deviations from the SM prediction can be caused by anomalous coupling with the $W$ boson, new gauge or charged Higgs bosons, the presence of massive neutrinos, etc~\cite{Bryman:2021teu,Krawczyk:2004na,Chun:2016hzs,Marquez:2022bpg}.

If the daughter lepton polarization is not measured, only four Michel parameters, $\rho$, $\eta$, $\xi$, and $\xi\delta$, are accessible for the experiment. Currently, they are measured with a precision of several percent~\cite{Workman:2022ynf}. The Belle experiment has already achieved statistical uncertainty of an order of $10^{-3}$, while systematics are dominated by the trigger efficiency at a level of $10^{-2}$~\cite{Epifanov:2017kly}. Using the Belle~II full integrated luminosity of 50 fb$^{-1}$, a statistical uncertainty of $10^{-4}$ can be achieved~\cite{Epifanov:2020elk}, and the main task will be to control systematics at a competitive level. 

\begin{figure}[b]
\centerline{\includegraphics[width=1\textwidth]{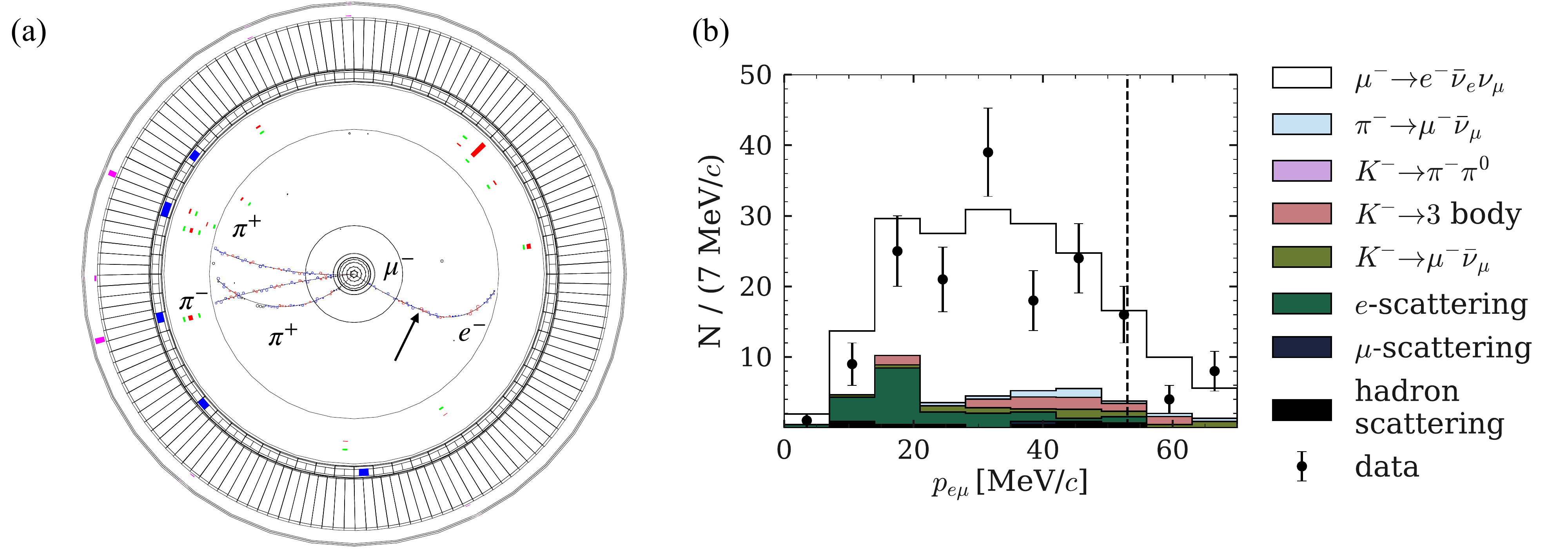}}
\caption{(a) Simulated event of $e^+e^-\to\tau^+\tau^-\to(\pi^+\pi^-\pi^+\bar{\nu}_\tau)(\mu^-\bar\nu_\mu\nu_\tau)$ with $\mu^-\to e^-\bar{\nu}_e\nu_\mu$ decay in the CDC (the arrow points to the decay vertex). (b) Distribution of the momentum of the daughter, in the rest frame of the mother, for electron and muon mass hypotheses assigned to the daughter and mother particles, respectively. The dashed line shows the $53\,\text{MeV}/c$ threshold. \label{fig:4}}
\end{figure}

The measurement of the daughter lepton polarization in the $\tau$ decay provides information on five more Michel parameters: $\xi^{\prime}$, $\xi^{\prime\prime}$, $\eta^{\prime\prime}$, $\alpha^{\prime}/A$, and $\beta^{\prime}/A$. The first one describes the longitudinal polarization of the daughter lepton. Using the full integrated luminosity of 988 fb$^{-1}$ collected by the Belle detector, we measure for the first time the Michel parameter $\xi^\prime$ in the $\tau^-\to\mu^-\bar{\nu}_\mu\nu_{\tau}$ decay to be $\xi^\prime=0.22\pm0.94\pm0.42$, where the first uncertainty is statistical, and the second one is systematic~\cite{Belle:2023udc,Belle:2023dyc}. We apply the innovative method based on the muon decay-in-flight reconstruction in the drift chamber as a kink~\cite{Bodrov:2021vkn,Bodrov:2021hfe} [the example of such event is shown in Fig.~\ref{fig:4}(a)]. 
The information about muon spin can be inferred from the daughter electron direction in the muon rest frame due to $P$-violation in the decay. The events were selected with a highly efficient machine learning algorithm trained to suppress the backgrounds from other kink events (pion and kaon decays, charged-particle scattering). The resulting distribution of the electron momentum in the muon rest frame for selected events is shown in Fig.~\ref{fig:4}(b).

The measurement uncertainty is dominated by the statistics, with systematics being under control with various data samples, including kaon and pion decay-in-flight from the $\tau$ and $D$-meson decays, electron scattering from $\gamma$-conversion, and hadron scattering from the $D$-meson decays. In addition to low statistics due to rare events of muon decays inside CDC, the sensitivity to the $\xi^\prime$ parameter is smeared by the absence of the special kink reconstruction algorithm. The development of such algorithm at Belle~II, together with enlarged CDC and record integrated luminosity of 50 fb$^{-1}$, can improve the statistical uncertainty up to $7\times10^{-3}$ with systematic uncertainty at the same level~\cite{Bodrov:2022mbd}.

\subsection{Radiative and five-body leptonic decays}
Radiative and five-body leptonic $\tau$ decays also provide information about the Michel parameters describing the daughter lepton polarization in $\tau$ decays~\cite{Shimizu:2017dpq, Flores-Tlalpa:2015vga}. In addition, their understanding is crucial for several LFV studies, where the radiative and five-body leptonic decays are the main backgrounds. 

Using the integrated luminosity of 711 fb$^{-1}$ collected by the Belle detector, we study the $\tau^-\to e^-\bar{\nu}_e\nu_\tau \gamma$ and $\tau^-\to \mu^-\bar{\nu}_\mu\nu_\tau \gamma$ decays and obtain the corresponding Michel parameters $\xi\kappa(e)=-0.4\pm1.2$, $\xi\kappa(\mu)=0.8\pm0.6$, and $\bar{\eta}(\mu)=-1.3\pm1.7$~\cite{Shimizu:2017dpq}. The introduced parameters are related to ones of the ordinary leptonic decays as $\xi\kappa=-1/4(\xi+\xi^\prime)+2/3\xi\delta$, and $\bar{\eta}=4/3\rho-1/4\xi^{\prime\prime}-3/4$.

Concerning the five-body leptonic decays, we conduct a feasibility study showing the possibility of reaching the branching fractions predicted in the SM for all modes with the integrated luminosity of 700 fb$^{-1}$ collected by the Belle detector~\cite{Sasaki:2017unu, Sasaki:2017msf}.

The studies described here can be repeated in the Belle~II experiment with higher precision.

\section{Search for $CP$ violation}
No CPV is observed in the charged leptons sector (in the SM, it is predicted only in the quarks sector). The most promising modes for the searches of CPV in the $\tau$ lepton decays are $\tau^-\to K^-\pi^0\nu_\tau$, $\tau^-\to K^0_S\pi^-\nu_\tau$, $\tau^-\to K^0_S\pi^-\pi^0\nu_\tau$, $\tau^-\to (\rho\pi)^-\nu_\tau$, $\tau^-\to (\omega\pi)^-\nu_\tau$, and $\tau^-\to (a_1\pi)^-\nu_\tau$. In the decay modes with $K^0_S$, non-zero CPV can be observed in the SM due to the presence of the effect in $K^0_S$ mesons.

The first measurement of the $CP$ asymmetry was performed by BaBar in the $\tau^-\to \pi^-K^0_S{\nu}_\tau$ decay:
\begin{equation}
    A_\tau=\dfrac{\Gamma(\tau^+\to \pi^+K^0_S\bar{\nu}_\tau)-\Gamma(\tau^-\to \pi^-K^0_S{\nu}_\tau)}{\Gamma(\tau^+\to \pi^+K^0_S\bar{\nu}_\tau)+\Gamma(\tau^-\to \pi^-K^0_S{\nu}_\tau)} = (-0.36\pm0.23\pm0.11)\%,
\end{equation}
where the first uncertainty is statistical, and the second one is systematic~\cite{BaBar:2011pij}. The obtained value is $2.8\sigma$ away from the SM prediction of $A^\text{SM}_\tau=(0.36\pm0.01)\%$.

It is also possible to use a modified asymmetry with differential distributions integrated over a limited volume in the phase space with a specially selected kernel, which is done in the Belle experiment~\cite{Belle:2011sna}. The obtained value of $A_\text{CP}$ is compatible with zero with the precision of around $10^{-3}$.

The most powerful method is to use unbinned maximum likelihood fit in the full phase space, which has not been done at $B$-factories so far. With this approach, using the full integrated luminosity, Belle~II can achieve a precision of $10^{-4}$~\cite{Kou:2018nap}.

\section{Search for lepton flavor violation}
Lepton flavor violating decays $\tau\to\ell\gamma$, $\tau\to\ell\ell\ell^{(\prime)}$, $\tau\to\ell h$ ($\ell,\,\ell^\prime={e,\,\mu}$, and $h$ is a hadron system), and modes with baryons in the final state are sensitive to new physics. In the SM, such decays are mostly forbidden, or their branching fractions are at the level of $10^{-53}$ and beyond any current or future experiment sensitivity, while different NP models predict them at the level of $10^{-7}$--$10^{-10}$. 

\begin{figure}[b]
\centerline{\includegraphics[width=0.9\textwidth]{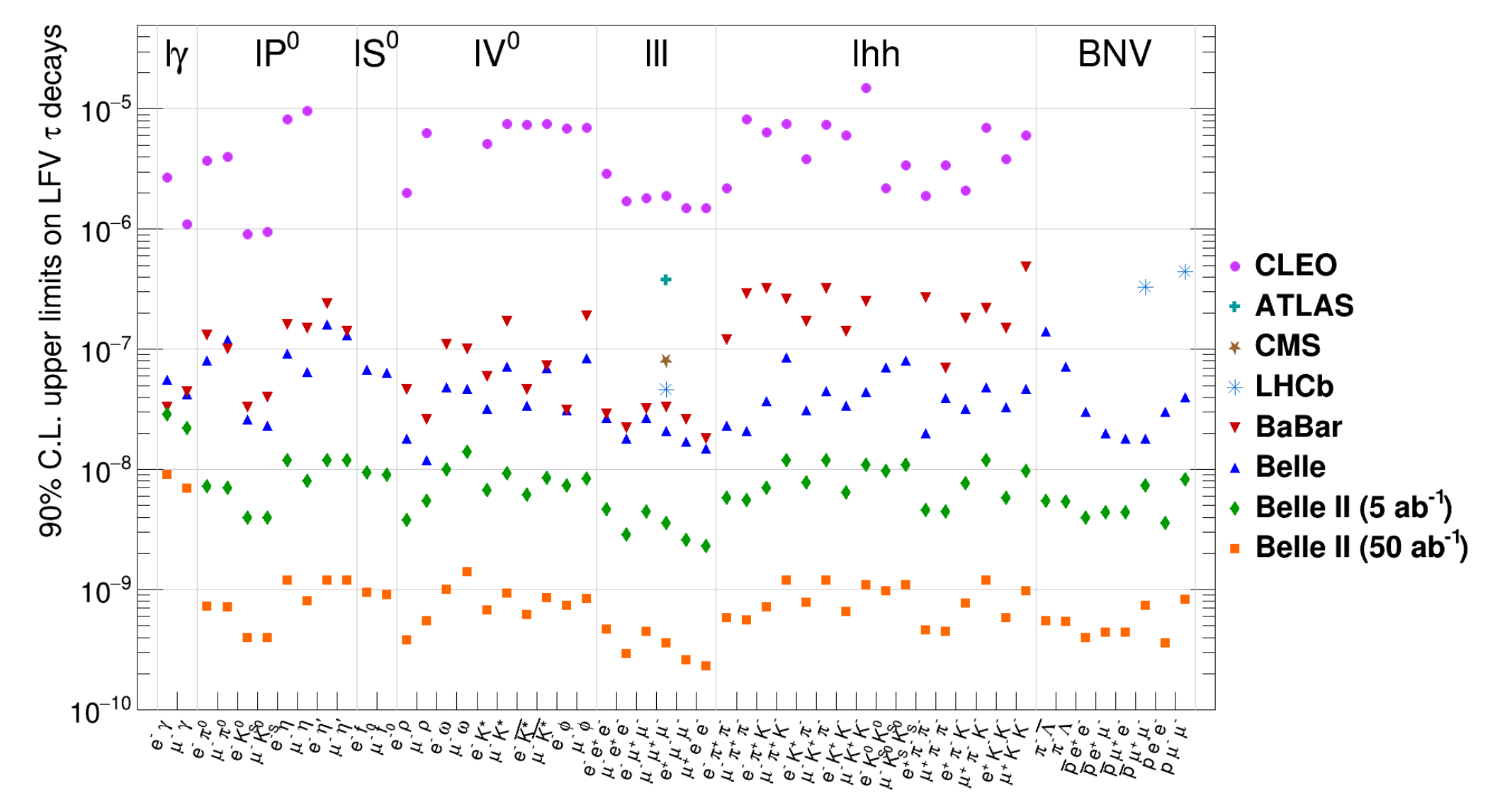}}
\caption{Projection of expected upper limits at the Belle II experiment~\cite{Banerjee:2022xuw} and current status of observed upper limits at CLEO, BaBar, Belle, ATLAS, CMS, and LHCb experiments~\cite{HFLAV:2022esi} on LFV $\tau$ decays. \label{fig:5}}
\end{figure}

The majority of the world's most stringent limits on these decays are obtained by the Belle collaboration, and they are at the level of $10^{-7}$--$10^{-8}$ (see Fig.~\ref{fig:5}). In the zero-background scenarios, Belle~II will improve Belle results linearly with the integrated luminosity increase, assuming the same analysis efficiency. The projections of the Belle precision with respect to an integrated luminosity of 5 ab$^{-1}$ and 50 ab$^{-1}$ for the Belle~II are shown in Fig.~\ref{fig:5}. For the $\tau\to\ell\gamma$ decay modes, there is an irreducible background leading to the limit improvement proportional to the square root of luminosity ratio only.

Although the Belle~II detector is still at the beginning of its data-taking, and the collected data sample is smaller than the one collected in the Belle experiment, Belle~II has already provided several new results in searching for lepton flavor violation in $\tau$ decays.

\subsection{$\tau^-\to\ell^-\phi$}
The first result of searching for LFV decays $\tau^-\to\ell^-\phi$ ($\ell = e$ or $\mu$) is provided by Belle~II using the integrated luminosity of 190 fb$^{-1}$~\cite{Belle-II:2023bnh}. For the first time, the selection of $e^+e^-\to\tau^+\tau^-$ events is based on an inclusive reconstruction of the non-signal $\tau$ decay (untagged approach), and the background is suppressed with a boosted decision tree classifier (BDT) application. To control the residual backgrounds, sidebands in data are used. Although the obtained boundaries of $\mathcal{B}(\tau^-\to e^-\phi)<23\times10^{-8}$ and $\mathcal{B}(\tau^-\to \mu^-\phi)<9.7\times10^{-8}$ (90\% CL for both) cannot compete with the previous Belle result of $\mathcal{B}(\tau^-\to e^-\phi)<2.0\times10^{-8}$ and $\mathcal{B}(\tau^-\to \mu^-\phi)<2.3\times10^{-8}$ (90\% CL for both)~\cite{Belle:2023ziz} due to the limited statistics, the final signal efficiency for the mode with the muon in the final state is two times improved compared to the former result.

\subsection{$\tau^-\to\mu^-\mu^+\mu^-$}
\begin{figure}[h]
\centerline{\includegraphics[width=0.6\textwidth]{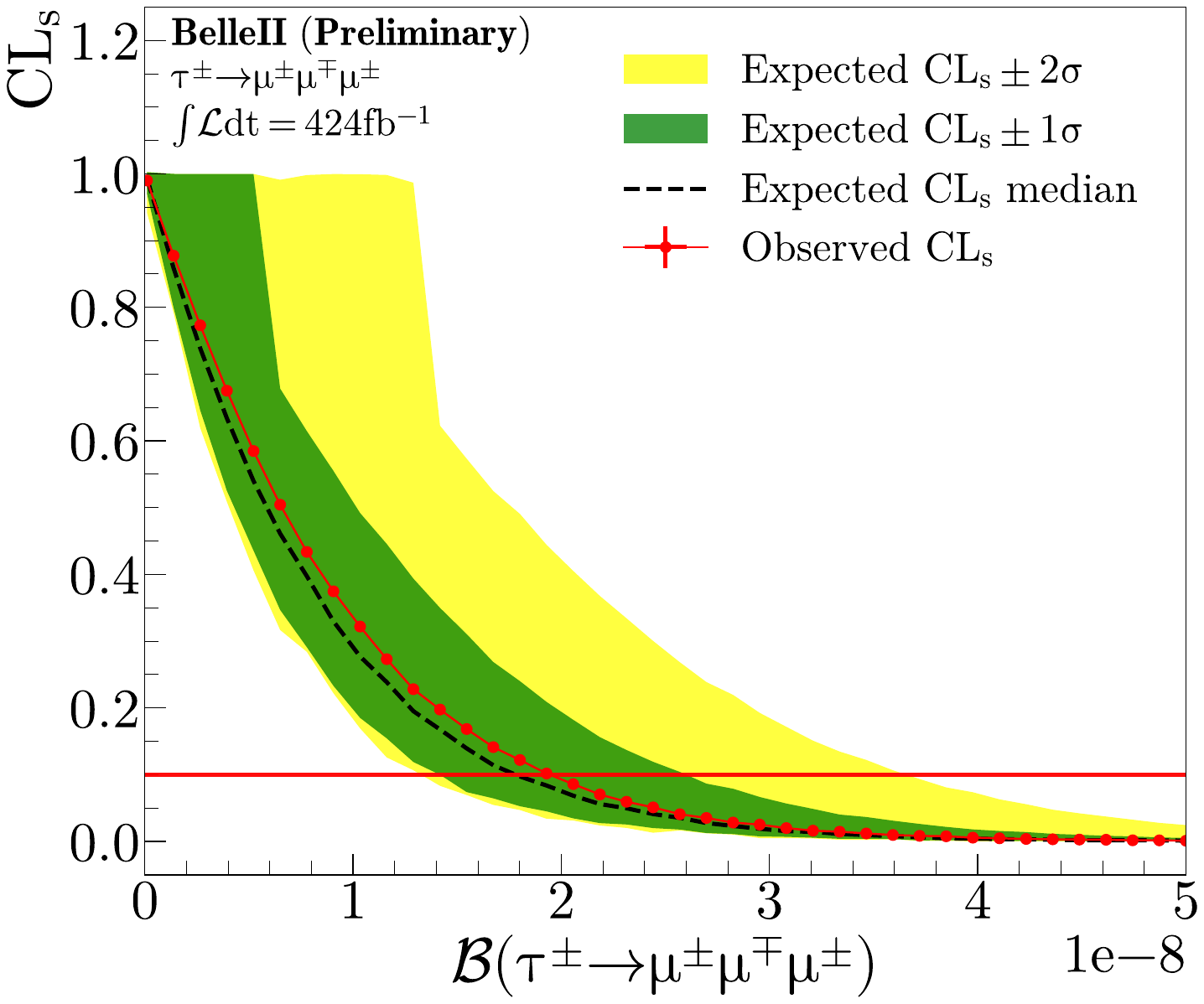}}
\caption{The plot with the upper limit on $\tau^-\to\mu^-\mu^+\mu^-$ decay. \label{fig:6}}
\end{figure}

The result of searching for $\tau^-\to\mu^-\mu^+\mu^-$ decay using the integrated luminosity of 424 fb$^{-1}$ collected by the Belle~II detector is already competitive with the one conducted by Belle using a 782 fb$^{-1}$ data sample~\cite{Belle-II:2024sce}. Again, the selection of $e^+e^-\to\tau^+\tau^-$ events is based on an inclusive reconstruction of the non-signal $\tau$ decay into a final state with one or three tracks and the BDT application. An alternative analysis using one-prong tagging reconstruction similar to that done previously by Belle and BaBar has been performed to validate the inclusive reconstruction and BDT approach. The final signal efficiency $\varepsilon_\text{sig}=(20.42\pm0.06)\%$ is almost three times better than the one obtained by Belle, retaining the same level of background. The plot with the upper limit is shown in Fig.~\ref{fig:6}. The 90\% CL limit on the branching fraction is $\mathcal{B}(\tau^-\to \mu^-\mu^+\mu^-)<1.9\times10^{-8}$, which is the most stringent boundary obtained up to now.

\subsection{LFV with invisible boson in final state}
\begin{figure}[h]
\centerline{\includegraphics[width=1\textwidth]{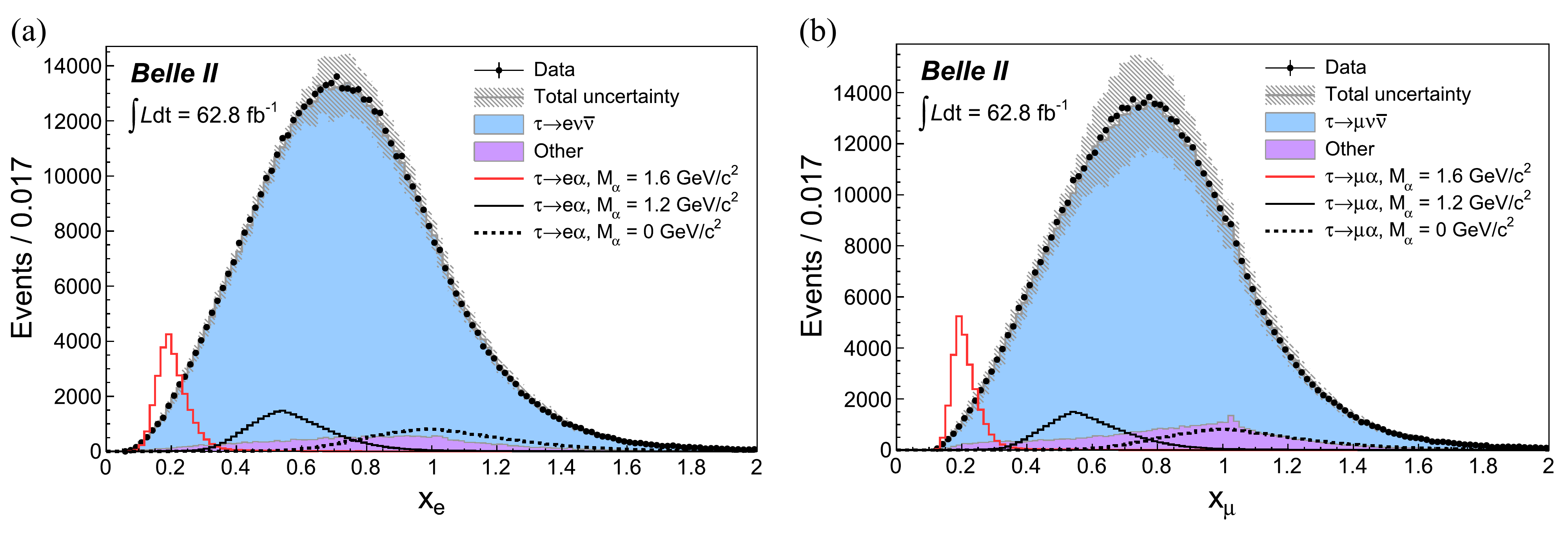}}
\caption{Spectra of $x_\ell$ for electrons (a) and for muons (b) in simulation and experimental data. \label{fig:7}}
\end{figure}

\begin{figure}[h]
\centerline{\includegraphics[width=1\textwidth]{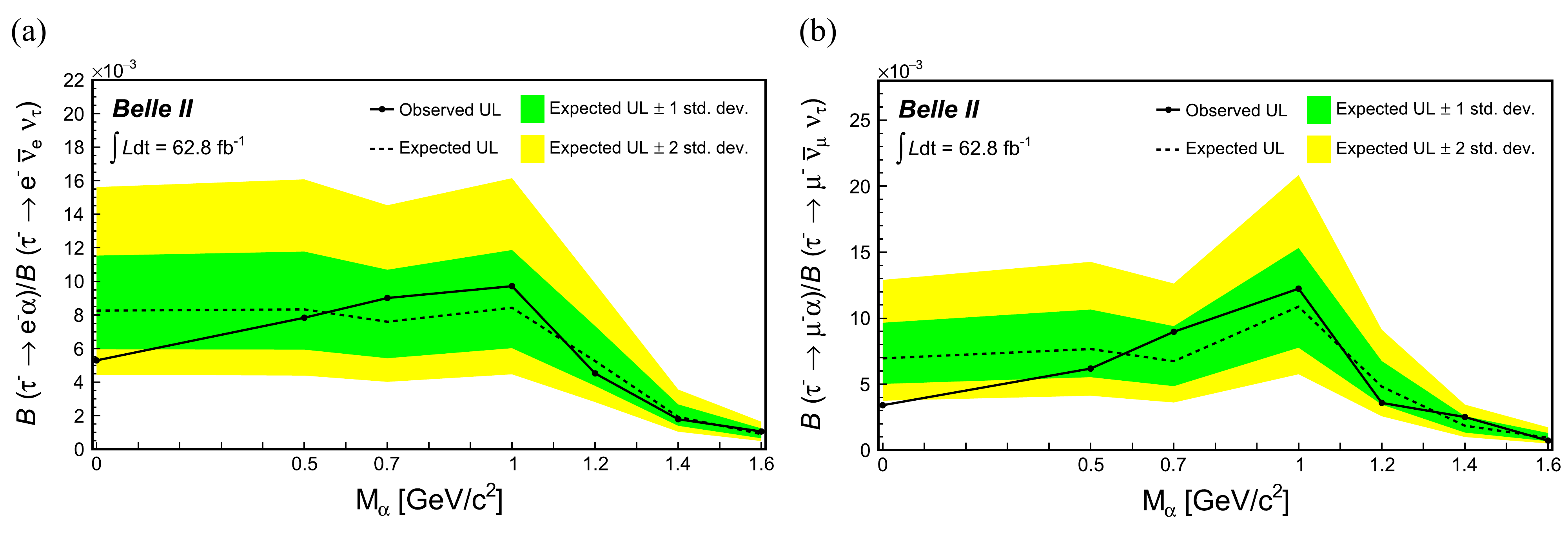}}
\caption{Upper limits at 95\% CL on the branching fraction ratios $\mathcal{B}(\tau^-\to e^-\alpha)/\mathcal{B}(\tau^-\to e^-\bar{\nu}_e\nu_\tau)$ (a) and $\mathcal{B}(\tau^-\to \mu^-\alpha)/\mathcal{B}(\tau^-\to \mu^-\bar{\nu}_\mu\nu_\tau)$ (b) as a function of the $\alpha$ mass. \label{fig:8}}
\end{figure}

Previously described LFV processes have only SM particles in the final state. It is also possible to search for decays with NP particles in the final state, like $\tau^-\to\ell^-\alpha$ ($\ell = e$ or $\mu$), where $\alpha$ is an invisible spin-0 boson predicted in models with axionlike particles. The study is performed using the integrated luminosity of 62.8 fb$^{-1}$ collected by the Belle~II detector~\cite{Belle-II:2022heu}. The non-signal $\tau$ lepton in the event is reconstructed in $\tau^+\to h^+h^-h^+\bar{\nu}_\tau$ decay mode ($h=\pi$ or $K$), and the momentum of the hadron system is used to build a pseudo rest frame for the signal $\tau$ lepton: $\vec{p}_\tau\sim-\vec{p}_{3h}/|\vec{p}_{3h}|$. The signal process is searched as excess above the $\tau^-\to\ell^- \bar\nu_\ell\nu_\tau$ spectrum in variable $x_\ell=2E_\ell/m_\tau$ (it is shown in Fig.~\ref{fig:7}). The resulting upper limits on the branching fractions of the $\tau^-\to\ell^-\alpha$ decays are shown in Fig.~\ref{fig:8}, depending on the mass of the invisible boson. The obtained result is from 2.2 to 14 times more stringent compared to the previous bounds provided by the ARGUS Collaboration~\cite{ARGUS:1995bjh}.

\section{Search for heavy neutrinos}
Heavy neutrinos or heavy neutral leptons (HNL) are introduced in different extensions of the SM to explain the generation of left-handed neutrino masses. We perform two searches for the HNL $N$ at the GeV mass scale produced in the $\tau^-\to\pi^- N$ decays using data collected by the Belle detector. In the first study, we assume $N$ to be a Majorana particle and reconstruct it decaying into $\pi^\pm\ell^\mp$ ($\ell=e$ or $\mu$) using 988 fb$^{-1}$ data sample~\cite{Belle:2022tfo}, while in the second one, we reconstruct $N\to\mu^+\mu^-\nu_\tau$ decay\footnote{Here, we assume that HNL mixes predominantly with $\nu_\tau$.} as a $\mu^+\mu^-$ displaced vertex ($>15$ cm from the beam axis) using 915 fb$^{-1}$ data sample~\cite{Belle:2024wyk}. In both analyses, we exclude from consideration the $K^0_S$ mass region. In the first case, no tag-side requirements are implied, and in the second case, we reconstruct one prong tag side.

\begin{figure}[h]
\centerline{\includegraphics[width=1\textwidth]{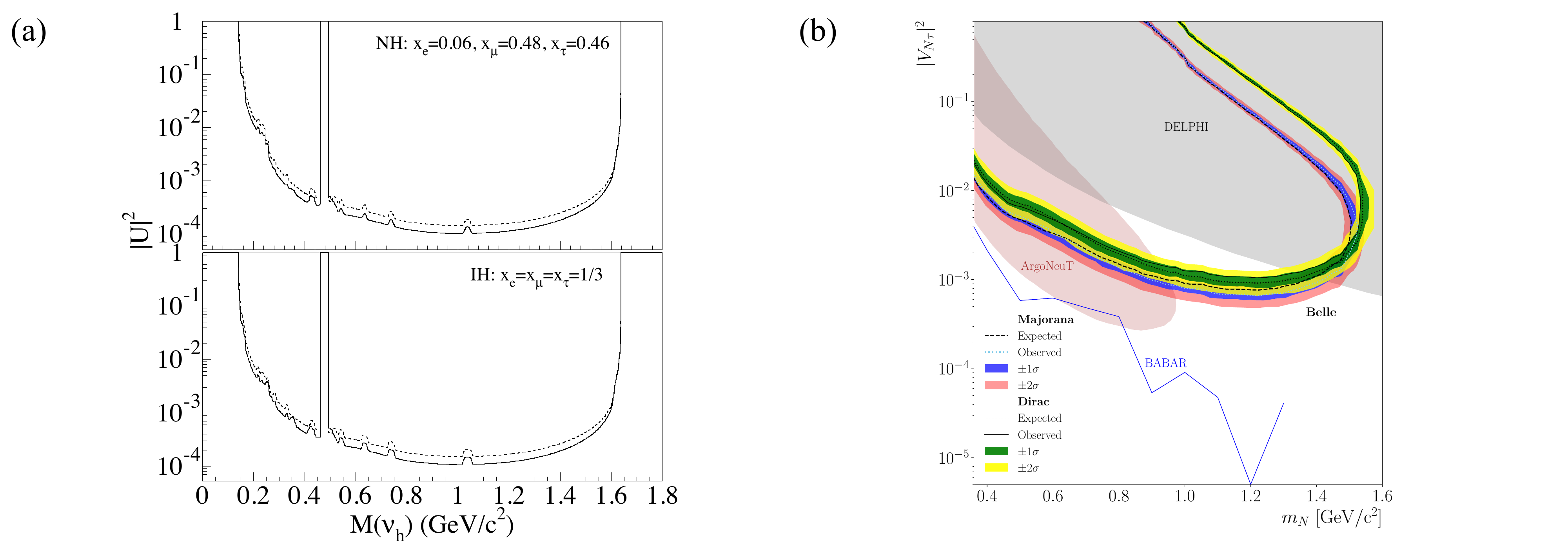}}
\caption{(a) Upper limits at 95\% CL on $|U|^2$ depending on the mass of the HNL $M(\nu_h)$. The upper (lower) plot is for the normal (inverted) hierarchy. (b) The expected (dashed) and observed (solid) 95\% CL limits on $|V_{N\tau}|^2$ depending on $m_N$ for a Dirac or Majorana HNL.
\label{fig:9}}
\end{figure}

We present the upper limits at 95\% CL on the sum of the mixing matrix elements $|U|^2$ in Fig.~\ref{fig:9}(a) for the first search and on the mixing coefficient $|V_{N\tau}|^2$ with $\nu_\tau$ in Fig.~\ref{fig:9}(b) for the second search. Both results are shown in dependence on the mass of the HNL.

\section{Conclusion}
The Belle experiment is the world's leading experiment in $\tau$ physics, providing a big part of the world’s best results. The Belle~II experiment with less than half the dataset of its predecessor has already provided competitive results and new methods applications: $\tau$ lepton mass measurement, search for LFV decays $\tau^-\to\ell^-\alpha$, $\tau^-\to\ell^-\phi$, and $\tau^-\to\mu^-\mu^+\mu^-$, and more are upcoming. In general, $\tau$ physics plays a significant role in the overall physics program of both the Belle and Belle~II experiments. 
It is worth mentioning that the Belle~II has a better sensitivity than the future Super Charm-Tau Factory~\cite{Charm-TauFactory:2013cnj, Luo:2018njj,Achasov:2023gey,Achasov:2024eua} in measurements that depend on the statistics, like most LFV searches, and in the measurement of the $\tau$ lepton lifetime that requires boost into the laboratory frame. 

By the end of the operation, the Belle~II experiment will accumulate an unprecedented number of $\tau^+\tau^-$ pairs, which makes it not only a Super $B$-factory but also a Super $\tau$-factory.

\section*{Acknowledgments}

The work is supported by the National Natural Science Foundation of China (NSFC) under Contracts No. 12335004. 
The work is prepared within the framework of the Basic Research Program at the National Research University Higher School of Economics (HSE).


\bibliographystyle{ws-ijmpa}
\bibliography{bibl}

\end{document}